\documentclass{article}

\PassOptionsToPackage{numbers, compress}{natbib}

\usepackage[final]{nips_2018}

\usepackage[utf8]{inputenc} 
\usepackage[T1]{fontenc}    
\usepackage{hyperref}       
\usepackage{url}            
\usepackage{booktabs}       
\usepackage{amsfonts}       
\usepackage{nicefrac}       
\usepackage{microtype}      

\usepackage{graphicx}
\usepackage{caption}
\usepackage{subcaption}
\usepackage{amsmath}
\usepackage{censor}

\usepackage{authblk}
\title{Computational EEG in Personalized Medicine:\newline A study in Parkinson's Disease}
\author[1]{Sebastian Mathias Keller}
\author[1]{Maxim Samarin}
\author[2]{Antonia Meyer}
\author[1,2]{Vitalii Kosak (Cozak)}
\author[2]{Ute Gschwandtner}
\author[2]{Peter Fuhr}
\author[1]{Volker Roth}
\affil[1]{University of Basel}
\affil[2]{Hospital of the University of Basel}
\date{}                     
\begin{document}

\maketitle

\begin{abstract}
Recordings of electrical brain activity carry information about a person's cognitive health. For recording EEG signals, a very common setting is for a subject to be at rest with its eyes closed. Analysis of these recordings often involve a dimensionality reduction step in which electrodes are grouped into 10 or more regions (depending on the number of electrodes available). Then an average over each group is taken which serves as a feature in subsequent evaluation. Currently, the most prominent features used in clinical practice are based on spectral power densities. In our work we consider a simplified grouping of electrodes into two regions only. In addition to spectral features we introduce a secondary, non-redundant view on brain activity through the lens of Tsallis Entropy $S_{q=2}$. We further take EEG measurements not only in an eyes closed (ec) but also in an eyes open (eo) state. For our cohort of healthy controls (HC) and individuals suffering from Parkinson's disease (PD), the question we are asking is the following: How well can one discriminate between HC and PD within this simplified, binary grouping? This question is motivated by the commercial availability of inexpensive and easy to use portable EEG devices. If enough information is retained in this binary grouping, then such simple devices could potentially be used as personal monitoring tools, as standard screening tools by general practitioners or as digital biomarkers for easy long term monitoring during neurological studies.

\end{abstract}

\section{Introduction}
\begin{figure}
\centering
  \includegraphics[width=.85\linewidth]{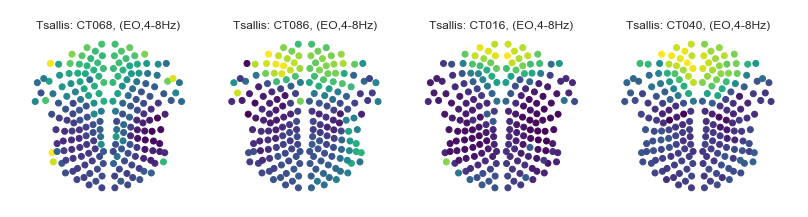}
  \caption{Estimated Tsallis entropy during eo state for the 4-8Hz band (averaged over several minutes) is shown for every electrode. Most but not all subjects of our cohort fall into this ``front versus back'' pattern where entropy is higher in the frontal brain regions. As subjects transition into ec state, frontal entropy is significantly reduced (not shown). CT068 and CT086 are healthy controls whereas CT016 and CT040 have been diagnosed with Parkinson's disease.}
\label{fig:tsEntropy}
\end{figure}

The recording of the first human EEG was performed in 1924 by German physician Hans Berger. Since then relative band power has become an established measure to quantify deviations from normal oscillatory brain activity: Several minutes of EEG signal, usually recorded under resting state condition and in an ‘eyes closed’ (ec) setting, are filtered into four to seven non-overlapping frequency bands covering a range from 0.5Hz up to 70Hz. Then, for each of those bands, the relative signal power is calculated. Based on this signal power, multiple studies have established a statistical connection between the neurological condition of their subjects (as measured e.g. by psychological testing) and the distribution of spectral power within these bands \citep{Dauwels}, \citep{Bon2008}, \citep{hardmeier2014reproducibility}.\\
High density EEG machines available today provide up to 256 individual electrodes recording brain activity at a frequency of 1000Hz or higher. If one is not interested in performing source reconstruction or spatial filtering such a high number of electrodes is usually not necessary, especially as electrodes in proximity to each other are very highly correlated. It is thus expected that features based on a grouping of electrodes into 10 or more regions as shown in fig. \ref{fig:reg10} are only slightly more informative as a grouping into only 2 such regions as shown in fig. \ref{fig:reg2}. In fact a color coded representation of the mean Tsallis entropy $\langle \hat{S}_{q=2}\rangle$ per electrode, as shown in fig. \ref{fig:tsEntropy}, indicates that the simplest viable partitioning is obtained by grouping electrodes into a frontal and an occipital region. If these two regions still carry enough information about a persons cognitive health it is potentially possible that two electrodes located at the front and back of the head would capture enough information to present a simplified alternative to a much denser electrode setting. Consequentially, as portable, longterm EEG devices with up to five electrodes are already available, these devices might prove useful as future long term monitoring tools analog to how ECG devices are used today to monitor heart rate variability.

\subsection{Parkinson's Disease and Dementia}
According to the American Psychiatric Association \citep{apa1994}, Dementia is a syndrome that consists of a decline in cognitive and intellectual abilities occurring in an awake and alert patient. As a symptom complex it is known to occur in over 70 disorders \citep{Geld1997}, among which Parkinson's Disease (PD) is the second most common age-related neurodegenerative disorder after Alzheimer's disease. It is estimated to affect nearly 2 percent of those over age 65. According to the National Parkinson Foundation (NPF), over the entire course of their illness, 50 to 80 percent of those suffering from Parkinson's disease eventually develop dementia. The average time from onset of movement problems to the development of dementia is about 10 years.
Accurately diagnosing Parkinson's Disease -- especially in its early stages -- requires experienced practitioners. Thus providing tools for detecting early changes in brain activity that are as easy to use as taking ones own blood pressure is an important first step.

\begin{figure}
\centering
  \includegraphics[width=.75\linewidth]{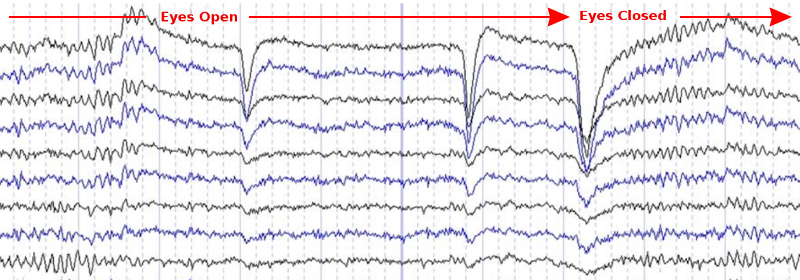}
  \caption{A significant change in brain oscillation can be observed when transitioning from an eyes open to an eyes closed state and vice versa. This change is known as the ``Berger effect'' or ``$\alpha$-blockade'' as the amplitude of EEG oscillations in the $\alpha$-band (8-13 Hz) decreases significantly when subjects open their eyes. Visible are also the three eye blinks during eyes open state.}
\label{fig:berger}
\end{figure}

\section{Materials and Methods}

In this study, we focus on how to complement and improve existing signal power based classification methods by (i) introducing additional entropy based features, (ii) including eo and ec states into the measurement process and (iii) using the Berger effect (see fig. \ref{fig:berger}) as a potential feature indicative of abnormal brain activity. At the same time we simplify the grouping of electrodes and investigate whether a measurement of frontal and occipital brain activity might potentially suffice to indicate a deviation from healthy brain oscillations.

\subsection{Sample Description}

We recruited participants for this study from the outpatient clinic for movement disorders of the Department of Neurology and Neurophysiology of the University Hospital of Basel (Universitätsspital Basel), Switzerland in the period from 2011 to 2012. We selected 46 persons with Parkinson's disease and labelled them as ``PD''. Parkinson's disease was diagnosed according to the United Kingdom Parkinson's Disease Society Brain Bank criteria. The patients who had dementia (Diagnostic and Statistical Manual of Mental Disorders, 4th Edition), history of stroke, epilepsy, multiple sclerosis and surgical interventions to the brain, or/and insufficient knowledge of German language, were excluded. We selected 16 persons without Parkinson's disease, matched to patients according to sex, age, and education, and labelled them as ``HC''. For controls, inclusion criteria were a subjective report of good health and a neuropsychological examination within normal limits. Exclusion criteria were identical to those for patients. The local ethics committee approved the study protocol. All participants provided written informed consent.

\subsection{Tsallis Entropy $\boldsymbol{S_{q=2}}$}
The q-entropy, also known as the Tsallis entropy, was proposed in \citep{Tsallis1988} and takes the following form
\begin{equation}
S_q(p_i) = \frac{1-\sum_{i=1}^{W}p_i^q}{q-1}
\label{eq:ts}
\end{equation}
with probabilities $p_i$ and $q\in\mathbb{R}$. In the limit $q \rightarrow 1$, eq. \ref{eq:ts} recovers the standard Shannon entropy. In analogy to the Normal distribution, which is the maximum entropy distribution of the Shannon entropy within the class of distributions with $E(x)=\mu$ and $E((x-\mu)^2)=\sigma^2$, Tsallis entropy $S_q$ is maximized by a q-Gaussian distribution. The pdf $p_{qg}$ of a q-Gaussian is given by 
\begin{equation}
p_{qg}(x) = p_0\left[1-(1-q)\frac{x}{x_0}^2\right]^{1/(1-q)},
\end{equation}
for details see section 3 of \citep{Picoli2009}. Fig. \ref{fig:q-Gaussian} shows the shape of such q-Gaussians for four different values of q where $q=1$ is the Normal distribution. Higher values of q lead to an increased probability in the tails of the distribution. Random samplings from a heavy tailed Gaussian with $q=2$ are shown in fig. \ref{fig:q-GaussianSamples} and compared to samples obtained for $q=1$. When band filtering a Gaussian white noise signal and a q-Gaussian ($q=2$) white noise signal we find that the signal obtained from the q-Gaussians has more similarities to a band filtered EEG signal for the same band in terms of amplitude variance than its Gaussian counterpart. This leads to the conclusion that Tsallis entropy might be a more appropriate entropy measure for EEG signals. This view is supported by several studies achieving improved classification results in similar settings based on Tsallis entropy, e.g. \citep{ref_ts2}, \citep{ref_ts1}, \citep{ref_ts3}. 

For estimating q-Tsallis entropy for $q=2$ an estimator $\hat{S}_{q=2}$ has been proposed in \citep{Sneddon2007}. It is parameter-free, easy to calculate, designed to estimate the entropy of time series and given by
\begin{equation}
\hat{S}_{q=2} = 1-\frac{\frac{1}{N}\sum s_i^2}{\sigma^2}
\end{equation}
with $s_i^2 = \frac{1}{n}\sum_{i=1}^n (x_i-\mu)^2$. To estimate the entropy of a time series it is first divided into $N$ bins where each extreme point of the signal marks the beginning, resp. the end, of a bin. The variance within each bin is given by $s_i^2$ whereas $\sigma^2$ is the variance of the entire time series.

\begin{figure}
\centering
\hspace{-1.6em}
\begin{subfigure}{.33\textwidth}
  \centering
  \includegraphics[width=1.0\linewidth]{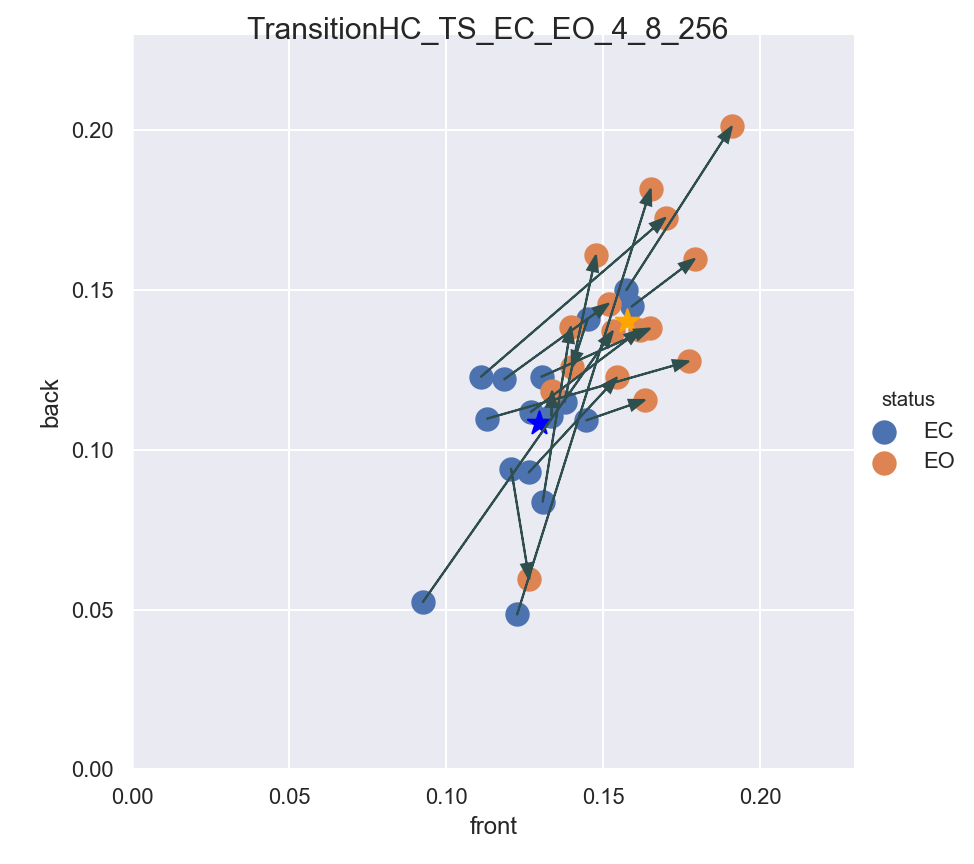}
  \caption{HC: ec $\rightarrow$ eo: for HC the occipital region has a higher increase in entropy in the eo state than for PD}
  \label{fig:sub1}
\end{subfigure}%
\hspace{.5em}
\begin{subfigure}{.33\textwidth}
  \centering
  \includegraphics[width=1.0\linewidth]{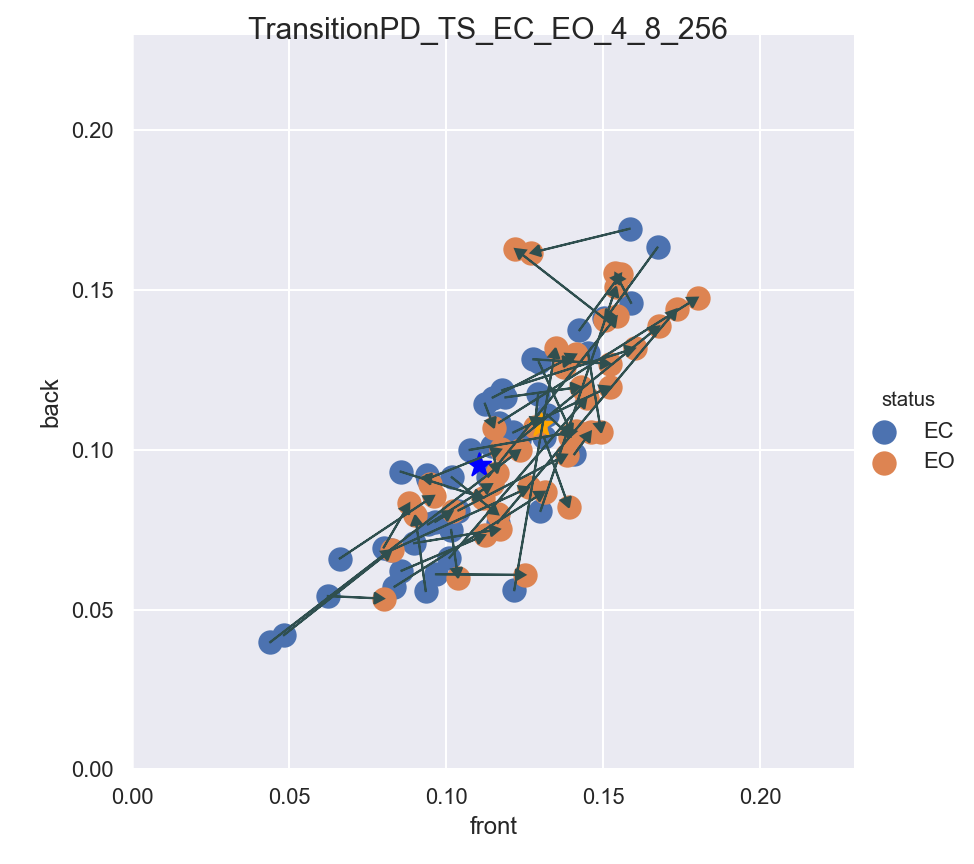}
  \caption{PD: ec $\rightarrow$ eo: for PD the increase in occipital entropy content is less pronounced}
  \label{fig:sub2}
\end{subfigure}
\hspace{.5em}
\begin{subfigure}{.33\textwidth}
  \centering
  \includegraphics[width=1.0\linewidth]{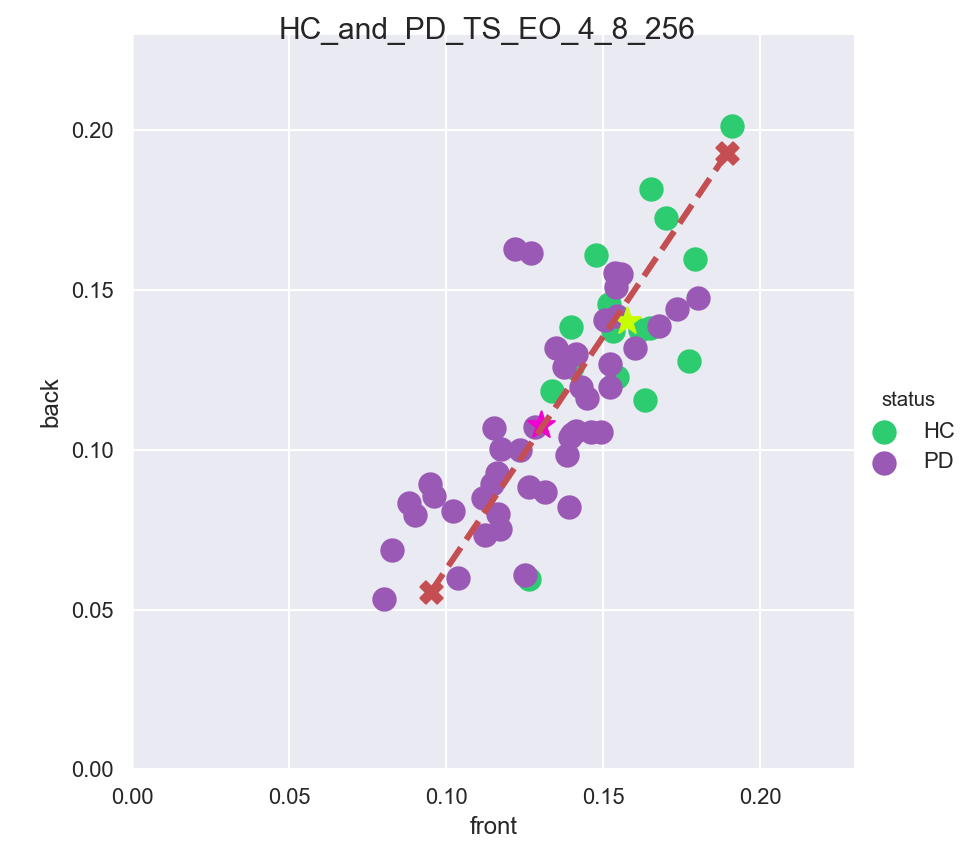}
  \caption{HC and PD group: the front to back ratio of entropy in the 4Hz-8Hz band is higher for healthy controls}
  \label{fig:sub2}
\end{subfigure}
\caption{ 4Hz-8Hz band: HC vs. PD in ec and eo state. In each panel the stars mark the means of the resp. groups. In panel (c) the crosses mark the extremes for HC and PD (during eo state) calculated using Archetype analysis \citep{Cutler1994}. They turn out to be the two most important features in table 1, row 4. Proximity to either extreme is an indicator of a person's neurological status, i.e. PD or HC.}
\label{fig:test}
\end{figure}

\section{Results and Discussion}

\begin{table}
  \caption{Binary classification with repeated cross validation (80/20) using gradient boosting machines (gbm) on a 2-region \textit{frontal vs. occipital} setting with target \textit{PD vs. HC}. BP is band power, TS is Tsallis Entropy. Measure is area under the ROC curve. Due to the imbalance between HC and PD group weighted classification and SMOTE were used but no significant difference in ROC was obtained.}
  \label{sample-table}
  \centering
  \begin{tabular}{lllll}
    \toprule
    \multicolumn{2}{c}{Classification (gbm)}                   \\
    \cmidrule(r){1-2}
    Feature Set     & Ratios & Modality & five most important Features (descending order) & ROC \\
    \midrule
    BP      & no  & EC      & bp.b.ec.4-8, bp.b.ec.13-45, bp.b.ec.1-4,   bp.f.ec.1-4,   bp.f.ec.4-8   & $64.0\%$ \\
    BP      & no  & EO      & bp.b.eo.1-4, bp.b.eo.4-8,   bp.b.eo.13-45, bp.f.eo.13-45, bp.f.eo.4-8   & $74.8\%$ \\
    TS      & no  & EC      & ts.f.ec.1-4, ts.f.ec.4-8,   ts.b.ec.8-13,  ts.b.ec.13-45, ts.b.ec.4-8   & $65.5\%$ \\
    TS      & no  & EO      & ts.b.eo.4-8, ts.f.eo.4-8,   ts.b.eo.13-45, ts.f.eo.8-13,  ts.b.eo.8-13  & $71.8\%$ \\
    BP + TS & no  & EC + EO & bp.b.eo.1-4, bp.f.ec.8-13,  ts.f.eo.4-8,   ts.b.ec.8-13,  ts.b.eo.1-4   & $\mathbf{75.7\%}$ \\
    Ratios(BP, TS) & yes    & EC + EO & bp.feo.vs.beo.1-4, ts.fec.vs.bec.8-13, bp.feo.vs.beo.4-8,     & $67.0\%$ \\
                   &        &         & ts.bec.vs.beo.13-45, bp.fec.bs.bec.1-4                        &          \\
    BP + TS & yes  & EC + EO & bp.b.eo.1-4, bp.feo.vs.beo.1-4, bp.feo.vs.beo.4-8, & $71.5\%$ \\
                   &        &         &   ts.fec.vs.bec.8-13, ts.f.eo.4-8                      &          \\

    \bottomrule
  \end{tabular}
\end{table}

EEG measurements in ec and eo states were filtered into four frequency bands (1-4Hz, 4-8Hz, 8-13Hz and 13-45Hz). For each band Tsallis entropy (ts) for $q=2$ and relative band power (bp) was calculated for the frontal (f) and occipital (b) region. PD vs. HC classification was performed with repeated cross validation on a train/test split of 80/20 using gradient boosting machines (GBM). Due to class imbalance the loss function was weighted accordingly and classification was measured using ROC. Our best result in the binary grouping setting is achieved with a combination of ec and eo measurements using entropy and band power features. Using only the ratios of band power and entropy in front and back gives only a moderate classification of ROC=67\%. Thus the absolute value of measured quantities in back and front are meaningful -- not only individual ratios. Combining ratios and absolute values of power and entropy does not improve overall classification: the possible benefit of additional features is outweighed by the larger search space. The largest increase in ROC was obtained when switching from ``eyes closed'' to ``eyes open'' measurements. We thus recommend to make it a standard procedure to record EEG measurements during \textit{both} ec and especially eo state. Our best classification result compares favourably -- given the reduced dimensionality -- with results based on 10 region groupings presented in \citep{Chat2017}, where ROC=80\% is achieved on a similar patient cohort also from the University Hospital Basel.

\subsubsection*{Acknowledgment} This project is supported by the the Swiss National Science Foundation project CR32I2-159682

\medskip

\small
\bibliographystyle{unsrtnat}

\bibliography{bibliography_eeg}
\newpage
\appendix
\section{Appendices}
\subsection{Simplified Partitioning: 10 regions vs. 2 regions}

\begin{figure}[ht!]
\centering
\begin{subfigure}{.5\textwidth}
  \centering
  \includegraphics[width=.60\linewidth]{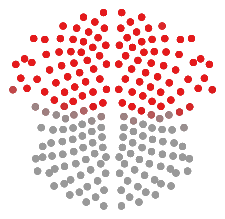}
  \caption{Electrodes partitioned into frontal and occipital\\region}
  \label{fig:reg2}
\end{subfigure}%
\begin{subfigure}{.5\textwidth}
  \vspace{-.6em}
  \centering
  \includegraphics[width=.8\linewidth]{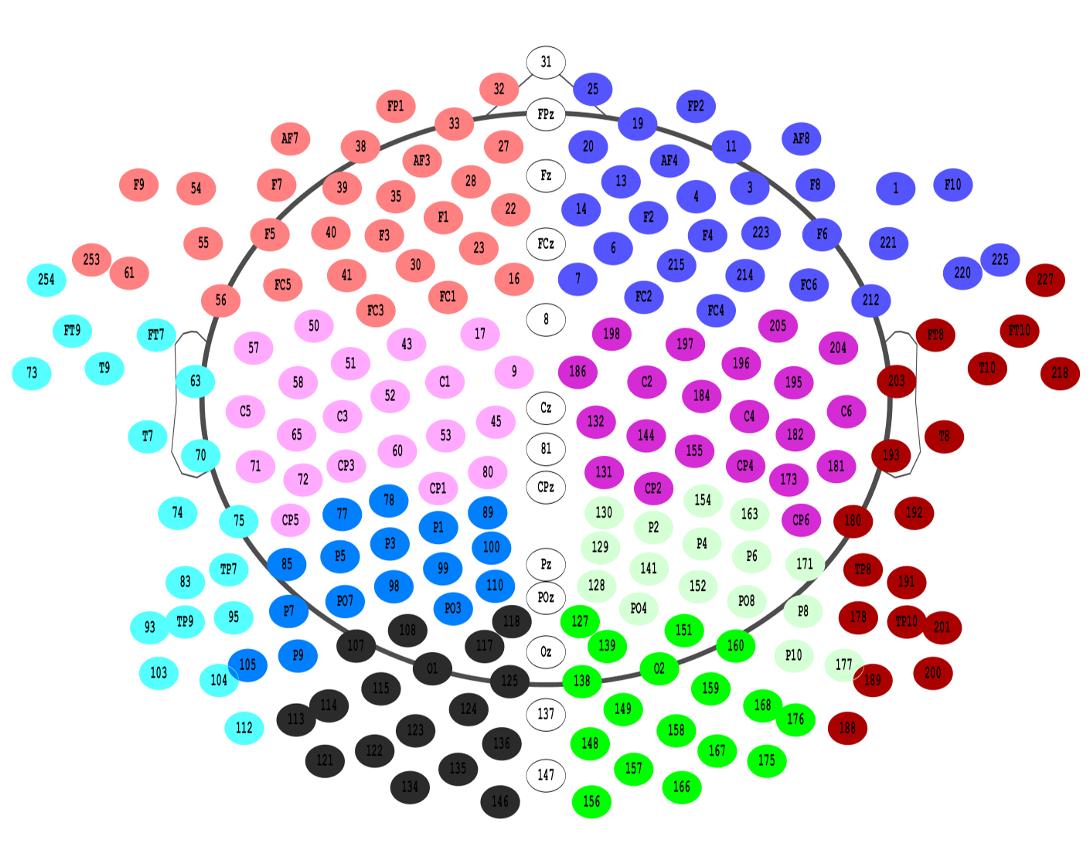}
  \caption{A standard partitioning into 10 brain regions as\\used in a clinical setting}
  \label{fig:reg10}
\end{subfigure}
\caption{The observed division into frontal and occipital region shown in fig. \ref{fig:tsEntropy} motivates the simplification of the standard partitioning with its 10 regions to a binary ``front vs. back'' partitioning of electrodes.}
\label{fig:test}
\end{figure}

\subsection{Family of q-Gaussians}

\begin{figure}[ht!]
\centering
\begin{subfigure}{.5\textwidth}
  \centering
  \includegraphics[width=1.05\linewidth]{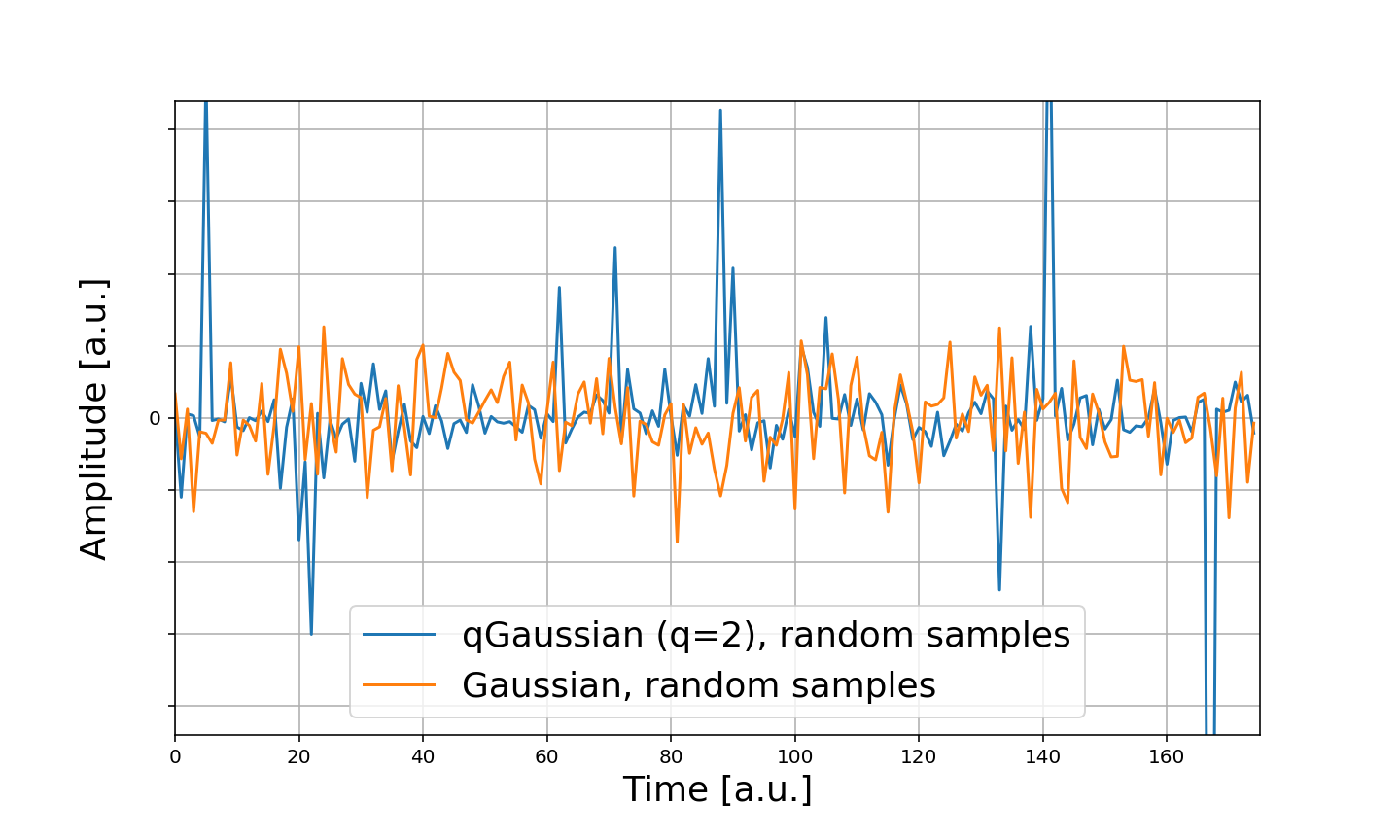}
  \caption{Random samples from a Normal distri-\\bution and from a heavy tailed q-Gaussian.}
  \label{fig:q-GaussianSamples}
\end{subfigure}%
\begin{subfigure}{.5\textwidth}
  \vspace{1em}
  \centering
  \includegraphics[width=1.05\linewidth]{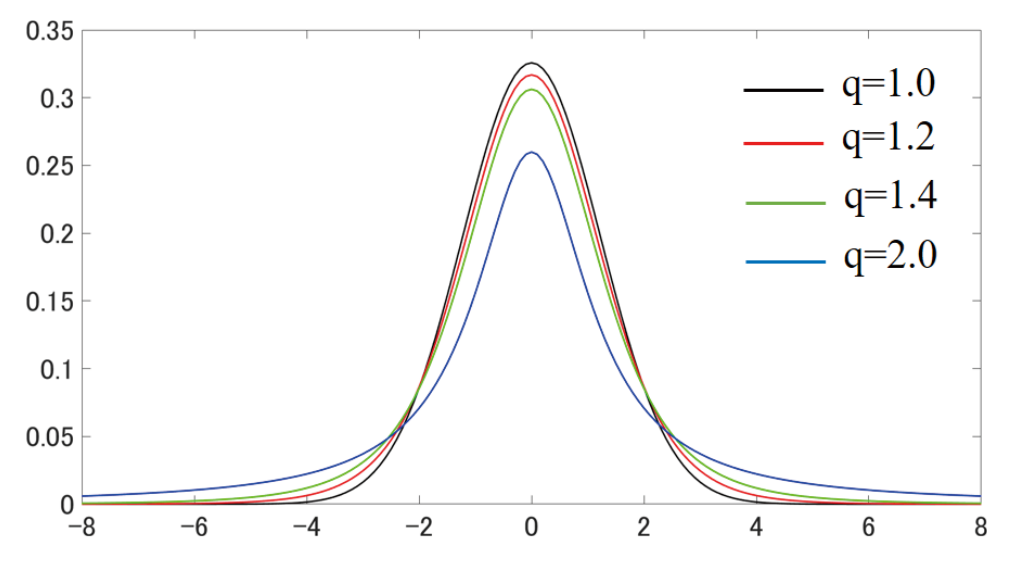}
  \caption{Within the family of q-Gaussian, the Normal distribution is recovered for $q=1$. Any q-Gaussian with $q>1$ has tails with higher probabilities.}
  \label{fig:q-Gaussian}
\end{subfigure}
\caption{In the family of q-Gaussians a distribution with lighter or heavier tails than the Normal distribution ($q=1$) is obtained for $q<1$ or $q>1$.}
\label{fig:test}
\end{figure}
\end{document}